\documentstyle[aps,manuscript]{revtex}
\begin{document}
\draft
\title{A Measurement of Gamow-Teller Strength for
$^{176}$Yb$\rightarrow$$^{176}$Lu and the Efficiency of a
Solar Neutrino Detector}
\author{M. Bhattacharya$^{1,2,3,a}$, C. D. Goodman$^1$, R. S. Raghavan$^4$, 
M. Palarczyk$^{5,b}$, A. Garc\'{\i}a$^2$,\\ 
J. Rapaport$^5$, I. J. van Heerden$^{1,c}$ and P. Zupranski$^6$}
\address{$^1$Indiana University Cyclotron Facility, Bloomington,
Indiana 47408, USA}
\address{$^2$University of Notre Dame, Notre Dame, Indiana 46556, USA}
\address{$^3$Nuclear Physics Laboratory, University of Washington, Seattle,
Washington 98195, USA}
\address{$^4$Bell Laboratories, Lucent Technologies, Murray Hill,
New Jersey 07974, USA}
\address{$^5$Ohio University, Athens, Ohio 45701, USA}
\address{$^6$The Andrzej Soltan Institute for Nuclear Studies, 
00-689 Warsaw, Poland}
\date{\today}
\maketitle
\begin{abstract}
We report a 0$^{\circ}$ $^{176}$Yb$(p,n)$$^{176}$Lu measurement at
IUCF where we used 120 and 160 MeV protons and the energy dependence
method to determine GT matrix elements relative to the Fermi
matrix element which can be calculated model independently.
The data show that there is an isolated concentration of GT strength 
in the low lying 1$^+$ states making the proposed Low Energy Neutrino 
Spectroscopy ({\em LENS}) detector (based on neutrino captures on 
$^{176}$Yb) sensitive to $^7$Be and {\em pp} neutrinos and a promising 
detector to resolve the solar neutrino problem. 
\end{abstract}
\renewcommand{\thefootnote}{\alph{footnote}}
\footnotetext[1]{E-mail:mbhattac@marie.npl.washington.edu}
\footnotetext[2]{Permanent address: Henryk Niewondnicza\'{n}ski 
Institute of Nuclear Physics, 31-342 Krak\'{o}w, Poland.}
\footnotetext[3]{Permanent address: University of Western Cape, 
Bellville, South Africa.}
\section{Motivation}
The existing solar neutrino detectors are sensitive to different
but overlapping regions of the solar neutrino spectrum. Combined 
data from these detectors have been used to estimate the individual
contributions from the $pp$, $^7$Be and $^8$B neutrinos to the
total solar neutrino flux. According to the present data the 
low energy $pp$ neutrinos seem to be present in full strength 
compared to the solar model prediction. The intermediate energy 
$^7$Be neutrinos seem to be missing entirely, while only about half 
of the high energy $^8$B neutrinos are observed. This energy dependent 
suppression of the solar neutrino spectrum is known as the 
modern {\em solar neutrino problem}.  There is a general 
consensus in the community that this energy dependent deficit of 
solar neutrinos cannot be explained by any reasonable 
modification of the solar model~\cite{ha:97}. The most 
plausible solution appears to be an energy dependent conversion 
of solar $\nu_e$'s to other flavors inside the Sun, {\em aka}, 
the MSW effect~\cite{ba:98,fi:99,fu:98,ab:99}. This idea can be 
fully tested only if the individual fluxes of at least $pp$ and 
$^7$Be neutrinos can be measured. 

The Low Energy Neutrino Spectroscopy ({\em LENS}) detector proposed 
recently by Raghavan~\cite{ra:97} will use a low-threshold, real-time, 
flavor-specific (using charged current neutrino capture) detection 
scheme based on Gamow-Teller (GT) transitions from 
$^{176}$Yb$\rightarrow$$^{176}$Lu for directly measuring the $^7$Be and 
the $pp$ neutrino flux. The $^7$Be $\nu_{e}$ flux measured by {\em LENS} 
along with the results from the {\em BOREXINO}~\cite{be:96} detector 
(sensitive to all flavors of $^7$Be neutrinos) will provide direct 
evidence of {\em Solar Neutrino Flavor Oscillation}. In addition 
{\em LENS} also promises to be the first detector capable of directly 
measuring the {\em pp} neutrino flux. 

In order to estimate the neutrino capture event rates for this detector 
one needs an accurate knowledge of the weak interaction matrix elements
for the transitions from the ground state of $^{176}$Yb to the two
low-lying levels in $^{176}$Lu. We report here a charge exchange
reaction [0$^\circ$ ($p,n$)] measurement of these matrix elements.

\section{Experiment}
At IUCF we measured $^{176}$Yb$(p,n)$$^{176}$Lu at 0$^\circ$ using the IUCF 
neutron time-of-flight ({\em NTOF}) system in spectroscopy mode, with 120 and 
160 MeV protons. We used a procedure, described in Refs.~\cite{ta:87,go:99}, 
in which the incident proton energy dependence of the ratio of the specific
GT to the specific Fermi cross section is exploited to determine from the
spectra the GT matrix elements relative to the model independent Fermi
matrix element. It is a well known fact that all of the Fermi strength
resides in the isobaric analog state transition, which is clearly seen
as a sharp peak in ($p,n$) spectra. The use of the energy dependence
method allows us to avoid the large uncertainties (as pointed out in 
Ref.~\cite{ta:87}) involved in using globally averaged reaction cross 
section on a single spectrum. Fig.~\ref{lowex-fig}(a) and (b) show the 
low energy part of the missing mass spectrum for 120 MeV and 160 MeV protons 
respectively. The data show that there is an isolated concentration of GT 
strength in the low-lying 1$^+$ states. 

The level scheme of $^{176}$Yb indicates that there are two 1$^+$ levels, 
not resolved in $(p,n)$ but resolved in a complementary ($^3$He,t) 
experiment at Osaka~\cite{fu:00}, in this region of concentrated GT 
strength. For {\em LENS} to be able to detect the {\em pp} neutrinos it 
is essential that the lower excitation energy state contain a significant 
portion of this GT strength. Although the two low lying states were not 
physically resolved in our measurement, it is possible to separate the 
GT strengths to the two states by fixing the peak shape parameters
to their accurately determined values from an auxiliary spectrum 
(described below) and by fixing the excitation energies of the two 
states to their well known values. Accurate peak shape knowledge is also 
needed to determine precisely the number of counts in the Fermi peak used 
in normalizing the spectra to B(GT).

\subsection{Peak shape Analysis}

Neutron time of flight peak shapes are determined by the response of 
the detection system and the convoluted energy-time profile of the
proton beam bunch. An additional effect that contributes to the 
peak shape is the scattering of the neutrons by the ground below and 
near the detectors. Our empirical peak shape consists of a Gaussian
(accounting for the detector response, the beam energy profile, and
the primary time structure of the beam bunch) convoluted with a fast 
exponential on either side of the Gaussian to fully take into account
the time structure of the beam bunch. To take into account the 
contribution to the peak shape by the neutrons scattered from the
ground below the detectors a long (slow) exponential was also convoluted
to the low neutron energy (high excitation energy) side of the 
Gaussian, as these neutrons traverse a longer path than the direct 
neutrons.  As described in more detail in Ref.~\cite{go:99}, the 
counts in this tail are not from the solid angle of the detector 
and should ideally be removed from the analysis. 

In this experiment, however, we make use only of the relative 
areas of the peaks corresponding to the low-lying final states 
and the IAS.  Using only relative peak areas eliminates 
systematic uncertainties involved in determining absolute cross sections. 
The uncertainty in a peak ratio resulting from an imperfect separation of 
direct peak from the tail is only second order due to a possible change 
in the direct to scattered ratio between the unknown GT transition and  
the IAS peak.   
 
As we simply ascribe a functional form to the observed peak shapes
rather than modeling them from first principles,  
it is extremely important to test our empirical peak shape and to  
obtain the best values for its free parameters from  an auxiliary 
spectrum using the same experimental setup. We chose the 
$^{13}$C$(p,n)$$^{13}$N reaction because of its well resolved level  
structure for this purpose. Our fit to the $^{13}$N  ground state 
using this peak shape and the Levenberg-Marquardt method is shown 
in Fig.~\ref{13clshap-fig}. The fit shows that our peak shape 
describes the data very well. We used the peak shape parameters 
obtained from this peak to fit the $^{176}$Lu spectrum shown in 
Fig.~\ref{lowex-fig}(a) and (b).

\subsection{Transition Strengths}
As described in detail in Refs.~\cite{ta:87} and~\cite{go:99}, 
the specific Fermi and GT cross sections have a strong mass number 
dependence that can not be accurately predicted by reaction dynamics 
theory. Hence the best way to normalize spectra to GT strengths is to 
normalize them to the Fermi transition. This procedure
requires obtaining the number of counts in the Fermi peak
which is usually not isolated or resolved from nearby and underlying
GT transitions. We use the strong incident proton energy 
dependence of the ratio of the Fermi to GT specific cross 
sections to extract the number of counts in the Fermi peak. In 
this procedure we match spectra taken at two proton energies at a 
GT transition close to the Fermi peak and iteratively subtract a 
peak of appropriate shape from the Fermi peak in each spectrum, 
while minimizing a figure of merit defined to be the sum of 
squared differences between the two spectra. The only assumptions 
that go into this procedure are that the nearby peak corresponds 
to a pure GT transition and that the energy dependence of the 
ratio of specific cross sections is universal ({\em i.e.} no mass 
dependence).

In this case (see Fig.~\ref{iasgtgr-fig}) the Fermi peak rides on top a 
smooth Gamow Teller Giant Resonance (GTGR). Although the GTGR does not 
have a smooth structure in lighter nuclei, one expects it to be so in 
heavier nuclei as in our case. As a result, in this case one can fit the 
entire region containing the Fermi peak and the GTGR using our empirical 
peak shape for the Fermi peak and approximating the GTGR shape by a 
Lorentzian function. Fig.~\ref{iasgtgr-fig}(a) and (b) show our best fit 
(${{\chi}^2_{\em min}}/\nu$=1.2) to this region for the 120 and 160 MeV 
spectrum respectively and we can see that this model describes
the data very well.

We separated the GT strengths to the two states in $^{176}$Lu by fixing
in our fit the parameters of our empirical peak shape to their values 
obtained from the $^{13}$N spectrum.  We also fix the well known excitation
energies of the two low lying states in $^{176}$Lu. The resulting fit 
(${\chi^2}/\nu$=1.1) is shown in Fig.~\ref{lowex-fig}. The GT strengths to the 
two states using the areas obtained from this fit and the area for the Fermi 
peak obtained from Fig.~\ref{iasgtgr-fig} is given in Table~\ref{gt-tab}.

\section{Neutrino absorption cross section on $^{176}$Yb}
In the allowed approximation, the neutrino absorption cross section on 
$^{176}$Yb is given by:
\begin{equation}
\sigma(E_\nu)=\frac{g_V^2 }{\pi \hbar^4 c^3} 
\sum_i p_i W_i F(Z,W_i)\left({g_A \over g_V}\right)^2B_i(GT)~. 
\end{equation}
The sum runs over all $^{176}$Lu daughter levels, $g_V$ and $g_A$ are the 
$\beta$-decay vector~\cite{sa:95} and axial vector coupling constants
respectively, $p_i$ and $W_i$ refer to the momentum and total energy of the 
outgoing electron, and $F(Z,W)$ accounts for the Coulomb distortion of the 
outgoing electron wave function. We computed $\sigma(E_{\nu})$ for the 
neutrino-capture 
reactions using the $B_i$-values from our measurement. We calculated 
$F(Z,W)$ using our codes for $F_0$ and screening corrections and by 
interpolating Behrens and J\"anecke's~\cite{be:69} $L_0$ value (their Table 
II) for finite-size corrections. Fig.~\ref{nuex-fig} shows the 
$^{176}$Yb $\nu$-absorption cross section as a function of the neutrino energy.

Our integrated $^{176}{\rm Yb}(\nu_e,e)^{176}{\rm Yb}^{\ast}$ cross 
section over the standard (no neutrino oscillations) {\em pp} 
and $^7$Be neutrino spectra is shown in Table~\ref{gt-tab}. LENS will 
also be a real time detector of supernova neutrinos and we obtain a total 
absorption cross section of $(5.57\pm 0.83)\times{10^{-42}}~{\rm cm}^2$ for 
$\nu_e$'s from a Fermi-Dirac energy distribution with T $=~3.2$ MeV 
(corresponding to $\langle E_{{\nu}_e} \rangle$ = 10 MeV).  


%
\section{Conclusions}
We measured the transition strengths of $^{176}$Yb$\rightarrow$$^{176}$Lu 
using 0$^{\circ}$ $(p,n)$ reactions. Our measurement indicates that this 
proposed detector will be sensitive to $^7$Be and $pp$ neutrinos making 
{\em LENS} a promising detector to help resolve the solar neutrino problem. 
The lowest state is sensitive to all solar neutrino sources including $pp$ 
neutrinos while the next excited state is sensitive to all sources except 
$pp$ neutrinos.  

\acknowledgments
We thank Bill Lozowski for preparing the rolled foil Yb target.
The IUCF and OU researchers were supported by the US National 
Science Foundation, (NSF). The Notre Dame researchers were 
supported by the NSF and the Warren Foundation.
%
\begin{table}
\caption{GT strengths and neutrino capture cross sections for the two low-lying states 
in $^{176}$Lu obtained from our spectra.}
\begin{tabular}{ccccc}
$E_x$(keV)~\protect\cite{nu:90} & B(GT) &
\multicolumn{3}{c}{$\sigma_{\nu}(10^{-45})~{\rm cm}^2$}\\
 &  &  $pp$ & $^7Be$(0.384) & $^7Be$(0.862) \\
\tableline
$194.511\pm0.008$ & $0.22\pm0.03$ & $11.16\pm1.53$ & $3.35\pm0.46$ & $77.95\pm10.69$ \\
$338.982\pm0.010$ & $0.12\pm0.02$ & - & - & $33.11\pm 5.75$\\
\end{tabular}
\label{gt-tab}
\end{table}
%
\begin{figure}
\caption{Low energy part of the $^{176}$Yb$(p,n)$$^{176}$Lu missing mass spectra, 
(a) using 120 MeV and (b) 160 MeV protons. Solid lines are composite fits to the 
spectra using the peak shape parameters obtained from $^{13}$N spectra. Dashed
lines show the individual peaks.}
\label{lowex-fig}
\end{figure}
\begin{figure}
\caption{Low energy part of the $^{13}$C$(p,n)$$^{13}$N missing mass spectra, (a)
using 120 MeV and (b) 160 MeV protons. Fit to the spectra using our empirical
peak shape is shown. Peak shape parameters obtained from these spectra are used 
in the subsequent fit to the $^{176}$Lu spectra.}
\label{13clshap-fig}
\end{figure}
\begin{figure}
\caption{High energy part of the $^{176}$Yb$(p,n)$$^{176}$Lu spectrum (a) using
120 MeV and (b) using 160 MeV protons. Overall fit to the spectra using the 
empirical peak shape for the IAS peak and a Lorentzian 
for the GTGR is shown.}
\label{iasgtgr-fig}
\end{figure}
\begin{figure}
\caption{Neutrino capture efficiency of LENS as a function of incident neutrino
energy.}
\label{nuex-fig}
\end{figure}
\end{document}